\documentclass[journal]{IEEEtran}
\usepackage{mathbbol}
\usepackage{amsthm,amsmath,amssymb}
\usepackage{dsfont}
\usepackage{mathrsfs}
\usepackage{amssymb,amsmath}
\usepackage{algorithmic}
\usepackage{algorithm}
\usepackage{url,subfigure,epsfig,graphicx}
\usepackage{float}
\usepackage{makecell}
\usepackage{CJKutf8}
\usepackage{stfloats}
\usepackage{graphicx}
\usepackage{subcaption}
\usepackage{color}
\ifCLASSOPTIONcompsoc

\usepackage{float}
\usepackage{makecell}
\usepackage[x11names,dvipsnames,table]{xcolor} 
\usepackage{colortbl}
\usepackage{multirow}
\usepackage{subcaption} 
\usepackage{lipsum}
\definecolor{mygray}{gray}{0.6}
\definecolor{myblue}{rgb}{0.8,0.85,1} 
\usepackage{array}
\newcolumntype{L}[1]{>{\raggedright\let\newline\\\arraybackslash\hspace{0pt}}m{#1}}
\newcolumntype{C}[1]{>{\centering\let\newline\\\arraybackslash\hspace{0pt}}m{#1}}
\newcolumntype{R}[1]{>{\raggedleft\let\newline\\\arraybackslash\hspace{0pt}}m{#1}}

\usepackage{array, tabularx, boldline}
\usepackage{eurosym}
\usepackage{amstext} 
\DeclareRobustCommand{\officialeuro}{%
	\ifmmode\expandafter\text\fi
	{\fontencoding{U}\fontfamily{eurosym}\selectfont e}}

  \usepackage[nocompress]{cite}
\else

  \usepackage{cite}
\fi

\ifCLASSINFOpdf

\else

\fi

\begin{document}

\begin{CJK}{UTF8}{gbsn}

\title{Toward Autonomous Digital Populations for Communication–Sensing–Computation Ecosystem}



\author{
\IEEEauthorblockN{Gaosheng Zhao and~Dong In Kim} \\
\IEEEauthorblockA{Department of Electrical and Computer Engineering, Sungkyunkwan University, Suwon 16419, South Korea}
\IEEEauthorblockA{\textbf{\emph{(Invited Paper)}}}
}

\maketitle

\begin{abstract}
Future communication networks are expected to achieve deep integration of communication, sensing, and computation, forming a tightly coupled and autonomously operating infrastructure system. However, current reliance on centralized control, static design, and human intervention continues to constrain the multidimensional evolution of network functions and applications, limiting adaptability and resilience in large-scale, layered, and complex environments.
To address these challenges, this paper proposes a nature-inspired architectural framework that leverages digital twin technology to organize connected devices at the edge into functional digital populations, while enabling the emergence of an evolvable digital ecosystem through multi-population integration at the cloud. We believe that this framework, which combines engineering methodologies with sociotechnical insights, lays the theoretical foundation for building next-generation communication networks with dynamic coordination, distributed decision-making, continuous adaptation, and evolutionary capabilities.
\end{abstract}


%



\section*{Introduction}

Communication networks have evolved significantly over the past decades, progressing from infrastructures focused solely on connectivity to intelligent systems that tightly integrate communication, sensing, and computation\cite{zhang2025integrated,wen2024survey,gonzalez2024integrated}. In current 6G visions, this integration is expected to deepen further, supported by enabling technologies such as space–air–ground architectures, intelligent reconfigurable surfaces, and large-scale edge–cloud collaboration\cite{siddiky2025comprehensive,dang2021should,zhu2024enabling}. These developments point toward a future where networks are not only connected and aware, but also adaptive, perceptive, and context-sensitive \cite{dong2024sensing,wild20236g}.

However, this functional convergence also introduces fundamental structural challenges \cite{ullah2025autonomous,cui2021inte,Ye2025energy,Stylianopoulos2025dis}. The integrated communication–sensing–computation architecture exhibits a pyramid structure: the lower layers consist of a vast number of devices equipped with three core functions, while the higher layers tend to concentrate on computing resources and retain only partial communication capabilities. The commonly adopted top-down centralized control and static hierarchical management approaches struggle to provide the adaptability required in large-scale, real-time environments. Moreover, the heavy reliance on human intervention further undermines the system’s autonomy.

To overcome these limitations, we propose a conceptual shift: to model future communication systems through the lens of natural ecosystems within the digital domain \cite{burattini2025distributing,he2025agent,brugiere2022handling}. Specifically, inspired by real-world ecological systems, we introduce the concept of digital populations at the edge, where multi-functional terminal agents are managed within the edge's computing capabilities and communication range. At the cloud level, multiple digital edge populations interact through distributed coordination, adaptive behavior, and feedback-driven evolution, collectively fostering the development of the whole digital ecosystem. Under this perspective, the network is no longer viewed as a mere collection of devices, but as an evolving society of intelligent agents governed through distributed mechanisms.

A key driver of this transformation is our introduction of a layered digital twin (DT) architecture that enables bidirectional interaction between the physical and digital domains. This architecture establishes the foundational abstraction of physical communication devices and environments into digital ecological replicas, encompassing local DTs that represent individual terminals, intermediate edge-level DTs that coordinate digital populations, and global DTs that manage strategy functions at the ecosystem scale. Obeying the hierarchical structure of current communication networks, the designed layered DT enables data to flow upward from the sensing input and supporting responsive actions at the multi-level.

Within the edge-level digital populations, communication, sensing, and computation are decoupled from the rigid boundaries of physical devices and reallocated as population-level functions. Local digital agents are dynamically assigned roles based on environmental conditions, device capabilities, and observed demands, with some agents specializing in data acquisition and upward transmission, and others focusing on local processing or coordination. Beyond the level of individual populations, we also propose an evolutionary approach at the cloud ecosystem scale. In this model, the behavioral rules and control structures of multi-edge populations are continuously reorganized and reconfigured in response to feedback signals, mobility dynamics, and environmental variability, while maintaining consistency under human-intervention functional goals.

In summary, this paper presents a layered DT framework that capitalizes on the inherent capabilities of communication, sensing, and computation to establish a self-organizing digital ecosystem. It describes how edge-level digital populations can be constructed on top of existing infrastructures, dynamically adapt to mobility and functional heterogeneity, and operate under the guidance of dual flows of information and authority to manage local environments autonomously, while remaining integrated with strategy, multi-population governance from the cloud.

\begin{figure*}[htbp]
	\centering
	\includegraphics[width=0.95\linewidth]{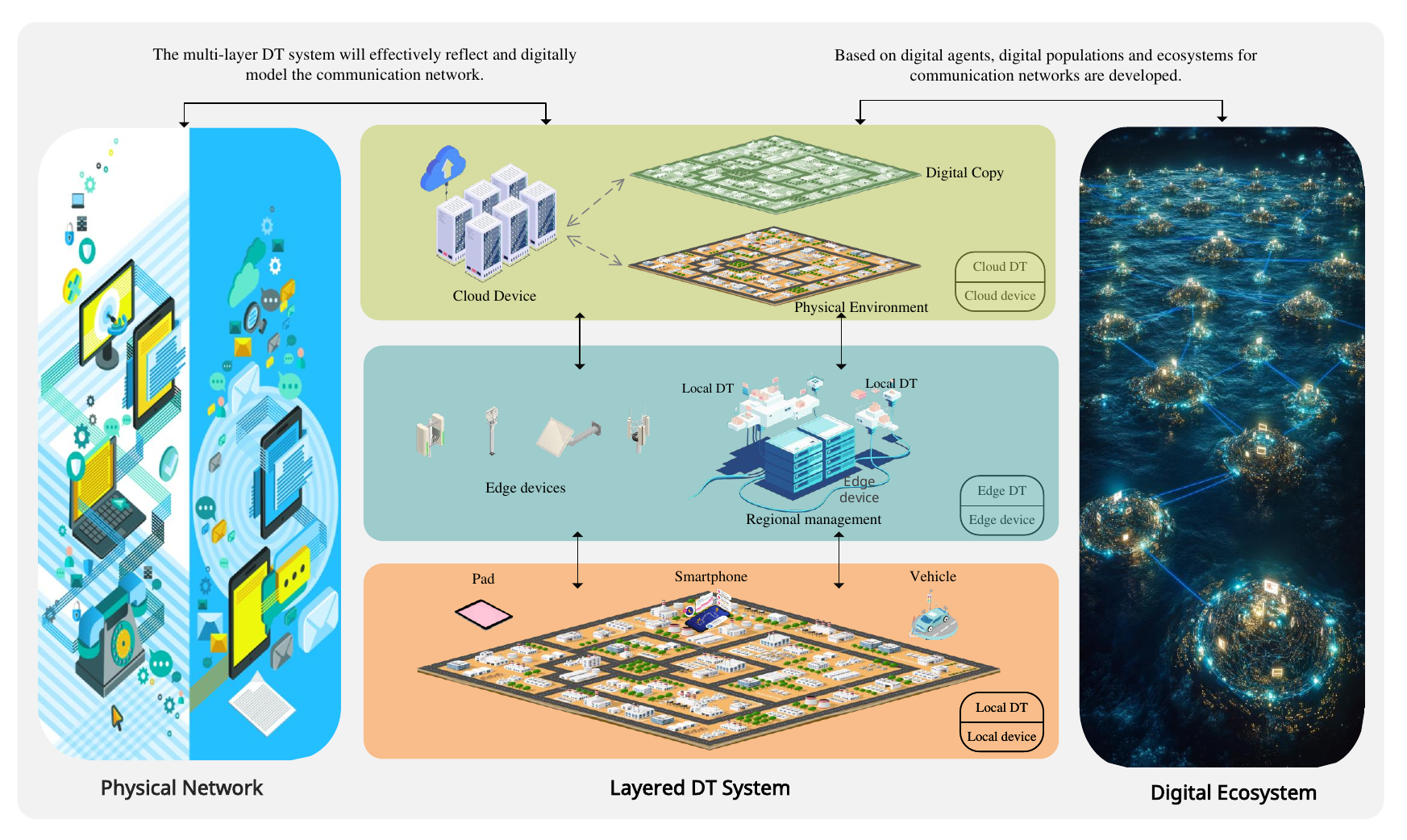}
	\caption{Progressive construction of a digital ecosystem: From physical networks to layered DT systems and finally to an evolving digital ecosystem.}
	\label{fig:layered-ecosystem}
\end{figure*}

To present this perspective in a coherent manner, the paper is structured to follow the progressive construction of a digital ecosystem, as illustrated in Fig.~\ref{fig:layered-ecosystem}. The system evolves from physical communication infrastructure to a layered DT system, and ultimately into an autonomous and adaptive digital ecosystem.

\section*{Step 1: Constructing an Interactive Replica of the Physical Environment — A Multi-Layer DT Framework}

The first step in abstracting a digital ecosystem from physical device systems is to construct an efficient and interactive digital modeling mechanism that enables real-time mirroring and intelligent management of the physical communication environment.

\subsection*{Digital Twins as the Intelligent Agents}

To realize this, DT technology is introduced as a core enabler. DTs maintain consistent and timely synchronization with their physical counterparts through continuous bidirectional data exchange\cite{zeb2022industrial,sharma2022digital,khan2022digital}. While DTs have already been widely applied in industrial and engineering domains for equipment monitoring, behavioral simulation, and decision support, their traditional use as static, human-facing replicas does not meet the demands of future communication systems.

Targeted communication systems are characterized by large scale, mobility, and functional integration of communication, sensing, and computation. As such, DTs should support these features systematically. Furthermore, DTs should not remain passive reflections. Instead, each DT must act as an autonomous digital agent that can interact persistently with its physical entity, perceive status and environmental changes, report operational feedback, and respond to high-level digital strategies.

In this context, DT modeling becomes more than just a mechanism for mapping physical states. It serves as the starting point for constructing autonomous and cooperative structures. Through this foundation, previously isolated physical nodes are integrated into a coordinated system that supports the formation of edge-level digital populations and enables long-term governance and evolution across the entire digital ecosystem.

\subsection*{Architecting the Twinning Pyramid: Local, Edge, and Cloud}

Modern communication networks are generally organized following a pyramid-shaped structure. At the base, a large number of resource-constrained devices carry out functions related to communication, sensing, and computation. These devices are diverse in distribution, often mobile, and primarily rely on wireless connectivity.  As a result, the local layer presents the greatest management complexity. Moving upward, the edge layer provides more concentrated functionality, utilizing moderate computational resources to analyze and control local devices within its communication domain. At the top, the cloud layer manages global services and policies, supported by powerful computing data centers and dedicated wired links.

To effectively map and utilize this hierarchical physical infrastructure, we introduce a multi-layer DT architecture composed of three closely connected levels: local, edge, and cloud. Each level maintains its own form of organizing DT representation, tailored to the functional and management needs in that layer.

\subsubsection{Local DT}

At the lowest level of the DT system, local DTs are embedded within various terminal devices such as sensors, smartphones, vehicles, and robots. Relying on each device’s limited computing and sensing capabilities, local DTs collect real-time information including internal states such as CPU usage and power consumption, external environmental conditions such as temperature, location, and channel quality, as well as user behavior data such as application usage patterns and mobility trajectories. After initial processing, this data is used to support local decisions, including executing lightweight inference models or temporarily switching communication strategies.

Much like biological individuals in natural ecosystems, each local DT operates autonomously within its environment, perceiving changes, reacting to local stimuli, and interacting with nearby entities. This autonomy enables local DTs to perform judgment and response directly at the terminal level, which helps reduce latency and alleviates the burden on upper layers. They also manage the upward transmission of processed data, task requests, and runtime status to edge DTs, while receiving policy feedback from higher layers. To enhance robustness in wireless environments, local DTs support horizontal communication with neighboring devices, facilitating basic collaboration and information sharing. 

Local DTs represent the most numerous, widely distributed, and frequently updated components of the digital ecosystem. They carry out micro-level responsibilities such as environmental perception, data preprocessing, and immediate response. Acting as both the sensory interface to the physical world and the initial node for upward information flow, each local DT lays the foundation for a modular, agent-based digital infrastructure.

\subsubsection{Edge DT}

At the middle layer of the digital ecosystem, the edge DT is deployed at edge nodes such as base stations and roadside units (RSUs). Its core task is to construct corresponding digital agents based on the data uploaded by local DTs within its coverage area. These agents are considered as digital individuals, collectively forming a regional digital population. Serving as the organizational and control center of this population, edge DT undertakes responsibilities such as state monitoring, resource scheduling, and strategy execution.

The primary characteristic of the edge DT is its ability for regional coordination. While local DTs focus on individual sensing and reaction, the edge DT abstracts and manages group-level behaviors. It integrates and allocates functions across multiple terminals within its domain, allowing the edge DT to balance the need for responsiveness with the complexity of coordination.

Another key feature is its structural flexibility and mobility awareness. The edge DT maintains a dynamic population structure where digital individuals can enter or exit the environment based on mobility patterns, resource availability, or communication conditions. This dynamic adaptation supports high-mobility and intermittently connected scenarios, similar to how populations reorganize in natural ecosystems. Through collaborative perception, distributed decision-making, and internal information exchange, the edge DT governs its local DT members.

In short, the edge DT functions as a mid-layer control entity within the digital ecosystem. It transforms distributed local DTs into a coordinated population, enabling efficient regional services and preparing the foundation for higher-level orchestration at the cloud layer.

\subsubsection{Cloud DT}

At the top layer, the cloud DT governs multi-population dynamics and acts as the global coordinator for long-term optimization, much like the invisible hand that guides balance in natural ecosystems. In contrast to the local and edge DTs that handle immediate sensing and regional coordination, the cloud DT functions as the strategy brain of the entire system. It continuously receives feedback from distributed edge DTs and integrates information across diverse geographic and functional domains.

The cloud DT enables the digital ecosystem to evolve over time. Through the analysis of historical records, behavioral trends, and inter-population interactions, it preserves an evolutionary memory that supports system-wide adaptation. This includes updating global models, refining coordination policies, and launching large-scale optimization strategies that reflect the long-term needs of the network.

From the ecosystem perspective, each edge DT is abstracted as a regional species embedded in a broader digital ecology. These populations may vary in density, mobility, service demands, and resource capabilities. The cloud DT is responsible for managing their coexistence and co-evolution, ensuring responsiveness, stability, and fair development across the system.

In summary, the cloud DT transforms scattered edge populations into a unified and evolvable digital ecosystem. By maintaining strategy foresight, adapting governance logic, and enabling scalable coordination, it constitutes the highest layer of intelligence within the multi-layer DT architecture.

\begin{figure*}[htbp]
	\centering
	\includegraphics[width=0.6\linewidth]{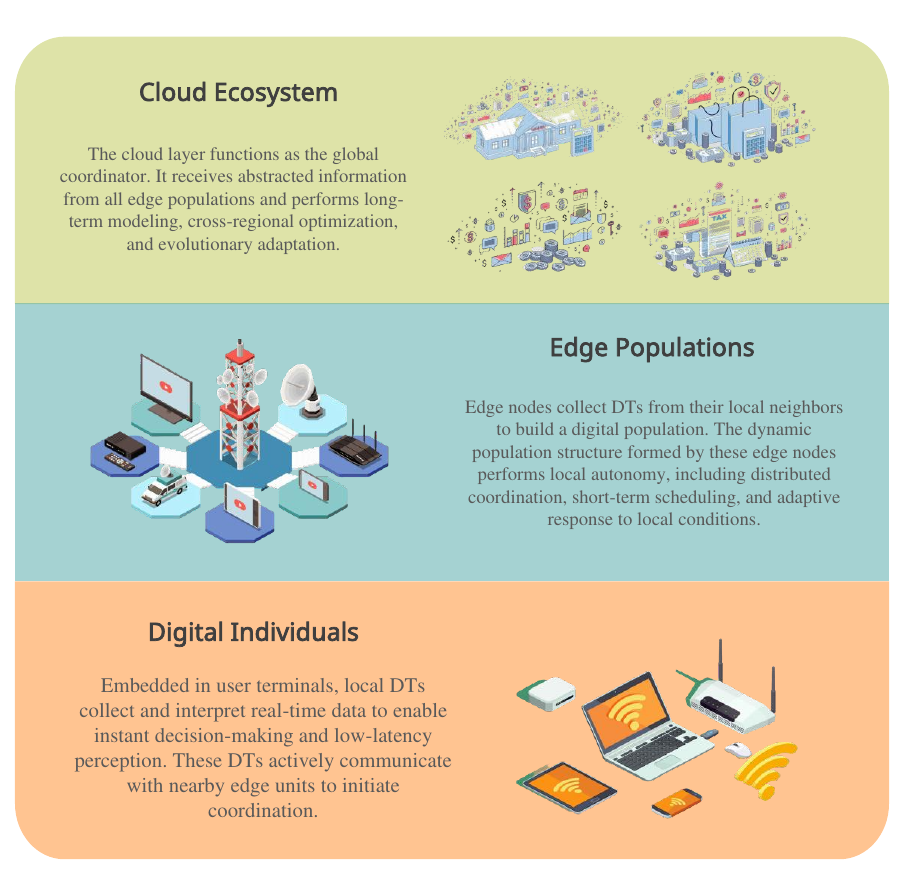}
	\caption{Digital ecosystem from multi-layer DT system.}
	\label{fig:layered-dt-structure}
\end{figure*}

As illustrated in Fig.~\ref{fig:layered-dt-structure}, the multi-layer DT system will evolve from physical communication infrastructures into a digital ecosystem composed of digital individuals, edge populations, and the cloud ecosystem. Each layer fulfills distinct yet collaborative roles: local DTs in user terminals enable real-time responsiveness, edge nodes coordinate distributed intelligence across mobile populations, and the cloud performs long-term modeling and ecosystem-wide evolution.

\section*{Step 2: Forming Mobile-Adaptive Digital Populations at the Edge DT Layer — Information Ascent and Strategy Descent}

Based on the “local to edge to cloud” structure of modern networks, the first step proposed a layered DT system architecture that lays the foundation for developing a digital ecosystem. In this framework, each device is treated as an individual entity equipped with sensing, communication, or computation capabilities. By binding with its DT, it becomes a digital unit that can be perceived, managed, and responded to within the edge layer.

Building on this foundation, the next step is to construct a dynamically evolving digital population at the edge layer. This population must adapt to continuous changes in the state and location of underlying devices, while also supporting the upward aggregation of local information and the downward dissemination of higher-level strategies. Functionally, this structure resembles middle ecological niches in nature. It mediates interactions among individuals and sustains adaptability at the population level in response to dynamic environments.

\subsection*{Edge-Level Digital Population Organization}

The primary task of the edge DT is to organize the originally scattered local DTs into a logically unified and dynamically adaptive digital population. Each local DT is mapped to a "digital individual" at the edge layer, continuously reporting its status and environmental data through sensing capabilities. Under the coordination of the edge DT, these individuals collectively form an expandable and contractible population.

Within this structure, the edge DT does not impose rigid centralized control over the digital individuals. Instead, it introduces a functional differentiation mechanism inspired by natural populations. Sensing, communication, and computation capabilities are decoupled from the fixed boundaries of physical devices and reallocated as population-level functions. Individuals are dynamically assigned different roles such as data acquisition and upward transmission, local data processing and aggregation, or intra-population coordination. The allocation of these roles depends on their capabilities, task loads, environmental conditions, and network status. 

Then, the dynamic organization of the population is reflected in two key aspects. First, structural flexibility: the edge DT adjusts its management boundaries in real time according to device location, link quality, or resource availability, enabling smooth entry and exit of terminal devices. Second, functional diversity: decoupled functions allow individuals to form a distributed division of labor across sensing, communication, and computation, improving adaptability to complex environments and supporting cooperative behaviors beyond the constraints of individual hardware capabilities.

\subsection*{Bottom-Up Information Ascent}

In the digital ecosystem, bottom-up information ascent is a core mechanism to support regional intelligence, vertical coordination and adaptation. The edge DT, functioning as the population control center, continuously receives heterogeneous data streams from its associated local individuals and builds a cohesive understanding of the regional environment.

This process begins with the on-device preprocessing capabilities of local individuals assigned sensing and computation roles. These agents perform operations such as downsampling, denoising, and feature extraction. The preprocessed information is labeled with device identifiers and contextual metadata, then transmitted by communication-role agents to the edge DT either periodically or in response to specific triggers.

At the edge DT, incoming data streams are aggregated and modeled to construct a structured view of regional states. This includes the fusion of sensing data such as environmental conditions and mobility, communication metrics such as link quality and congestion, and computational information such as task load and resource availability. Through spatial and temporal analysis, the edge DT identifies patterns, correlations, and latent risks across the digital individuals. For example, in urban scenarios, traffic flow maps can be generated in real time by aggregating speed and location data from multiple vehicle DTs, allowing the system to detect congestion or anomalies.

Processed insights are encapsulated into structured uplink packages and transmitted to the cloud ecosystem. Each package contains event labels, structured descriptions of current states, and trend indicators that support higher-layer strategy modeling, anomaly prediction, and policy refinement.

\subsection*{Top-Down Strategy Descent}

In the digital ecosystem, the edge DT not only supports bottom-up information ascent but also plays a central role in top-down strategy descent. The cloud ecosystem, based on cross-regional historical data and system-wide optimization goals, formulates strategies such as resource allocation policies, model update plans, or service-level adjustments. However, these strategies often require contextual refinement before they can be applied effectively to heterogeneous and dynamically changing edge populations. The edge DT acts as both a translator and scheduler of these strategies.

Specially, the edge DT interprets the strategy intent behind cloud-issued policies and adapts them to its current environmental context. Based on the available resources, it transforms high-level commands into executable actions suitable for each local individual. This includes dynamically reassigning sensing, communication, and computation roles to match the current operational priorities. For example, in scenarios with communication congestion, certain agents may be designated as high-bandwidth relays. In areas with degraded situational awareness, high-precision sensing nodes may be activated to improve environmental awareness.

As the edge DT advances in its ability to interpret and enact cloud strategies, it progressively attains higher levels of edge autonomy, thereby supporting decentralized coordination. It also forms the operational backbone for top-down guidance within the digital ecosystem, ensuring that the intentions of the cloud ecosystem are effectively realized across diverse and mobile edge populations.

\section*{Step 3: Cloud-Integrated Multi-Population Ecosystem — Evolutionary Population Governance}

Building upon the local digital replicas and the edge-level digital populations, the third step introduces a long-term governance mechanism at the cloud layer, aiming to establish an evolutionary multi-population ecosystem. Rather than relying on centralized control, this ecosystem leverages the cloud DT’s capabilities in historical accumulation, global awareness, and strategy planning to coordinate the co-evolution of multiple edge populations.

Serving as the highest-level intelligent core, the cloud assumes three essential roles. First, it acts as an integrator of knowledge by continuously receiving summarized information from edge populations and constructing a global understanding of the system state. Second, it functions as a strategy generator, creating actionable policies for lower layers based on historical behavior trajectories, trend predictions, and optimization objectives. Third, it serves as the system’s evolutionary memory, transforming long-term operational experiences, adaptation feedback, and dynamic resource distributions into experiential rules.

\subsection*{Knowledge Accumulation}

The cloud continuously aggregates uplink data packets from multiple edge regions to construct a cross-regional and cross-modal state graph. This graph not only includes raw indicators such as environmental measurements, resource metrics, and task distributions, but also integrates higher-level abstractions like behavioral patterns, service request trends, and the evolution paths of abnormal events.

This knowledge accumulation process parallels the concept of cultural memory or genetic inheritance in natural ecosystems. In biological populations, complex behaviors such as migratory routes or cooperative hunting are rarely the product of isolated individuals. Instead, they emerge through collective experiences, reinforced across generations. Likewise, the cloud layer of a digital ecosystem structures and models accumulated operational memory, refining its understanding of system dynamics and encoding this learning into strategy templates and predictive models.

\subsection*{Strategy Generation}

With its global view and historical memory, the cloud DT formulates a multi-population and cross-regional strategy framework. These generated strategies span across domains such as resource reallocation, task offloading, energy management, security enforcement, and model evolution.

Crucially, the cloud does not directly command execution at lower layers. Instead, it generates abstract policies tailored to specific regions. For example, based on predicted communication bottlenecks, the cloud might derive adaptive threshold rules and send them to edge DTs as parameterized policy blueprints. Each edge DT further localizes these policies, translating them into scheduling decisions, device activation plans, or content caching instructions.

This process resembles seasonal migration in animal populations. Environmental cues such as temperature shifts, day length changes, or resource availability act as large-scale signals that shape the overall movement and behavior of the population. While these signals provide a shared directional trend, each local group or individual adjusts its timing and route according to its own conditions and surrounding environment. 

\subsection*{Multi-Population Coordination}

Edge DTs independently manage their digital populations within bounded geographic or functional domains. However, for a digital ecosystem to maintain long-term adaptability and resilience, cross-population coordination is essential. The cloud DT achieves this by establishing mapping relationships among populations, identifying complementary pairs such as overloaded versus underutilized zones, and enabling resource balancing across the network.

For instance, when a specific region experiences computation overload, the cloud may coordinate low-load edge DTs from neighboring areas to contribute processing support.

\subsection*{Evolutionary Governance}

Ultimately, the goal of the cloud DT is to enable a self-improving and evolution-capable twin ecosystem. Through iterative strategy refinement, model retraining, and feedback incorporation, the system adapts to long-term trends while maintaining short-term responsiveness.

Rather than imposing fixed rules, the cloud supports gradual adaptation, similar to how natural evolution fine-tunes species through continuous interaction with the environment. Digital populations evolve through orchestrated simulations, controlled rollouts, and continuous monitoring.

In this way, the cloud DT transcends the role of a static controller. It becomes a meta-organizing agent, maintaining institutional memory, orchestrating high-level coordination, and steering the ecosystem’s long-term trajectory without disrupting localized autonomy.

\begin{figure*}[htbp]
	\centering
	\includegraphics[width=0.9\textwidth]{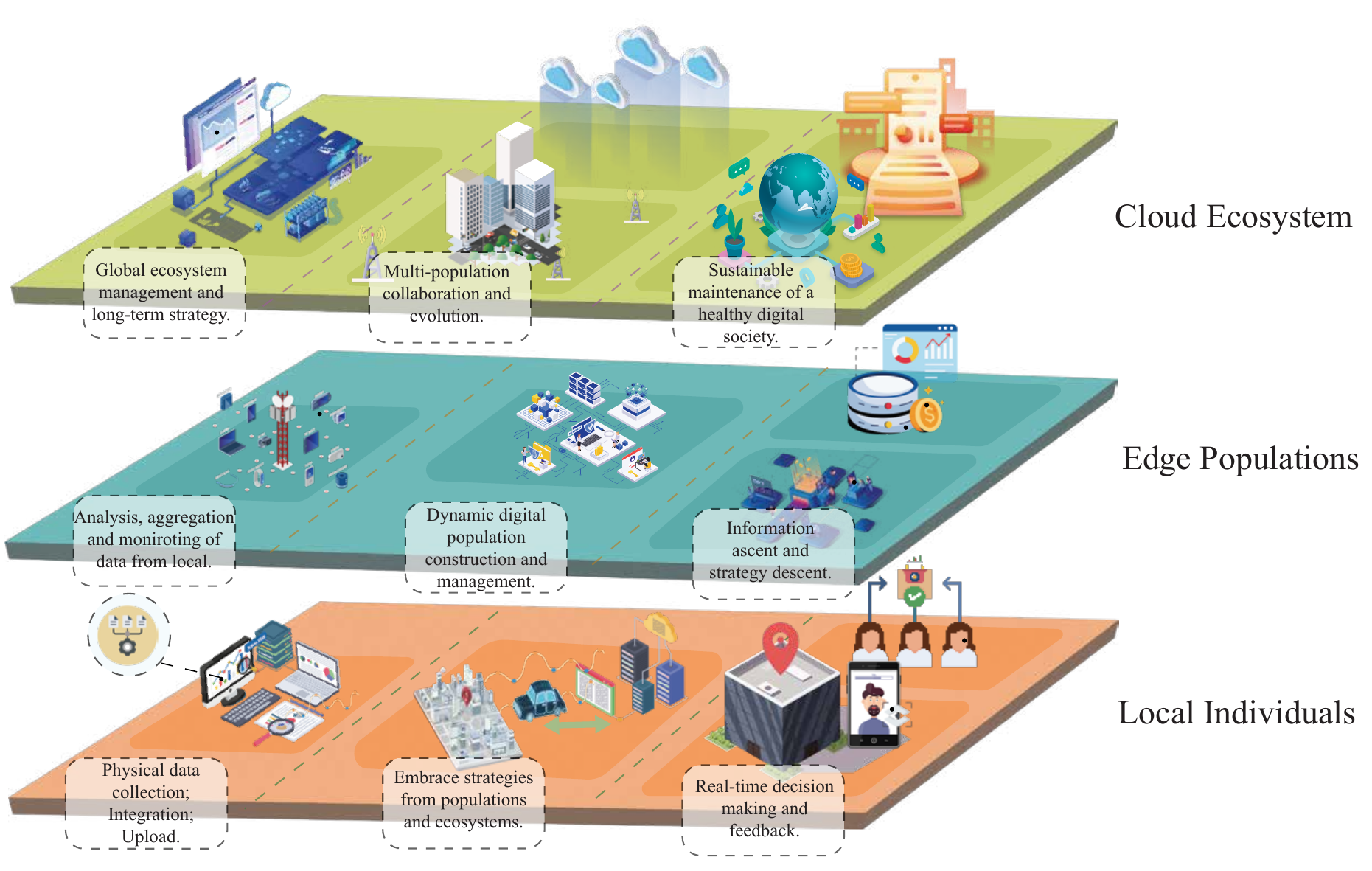}
	\caption{Digital ecosystem architecture composed of local individuals, edge populations, and cloud ecosystem, reflecting roles in real-time interaction, distributed coordination, and long-term governance.}
	\label{fig:3}
\end{figure*}

\section*{Step 4: A Healthy Digital Society – The Synergistic Coupling of Social and Natural Sciences}

\begin{figure*}[t!]
	\centering
	\subfigure[Response time distribution comparison.]{
		\includegraphics[width=0.4\textwidth]{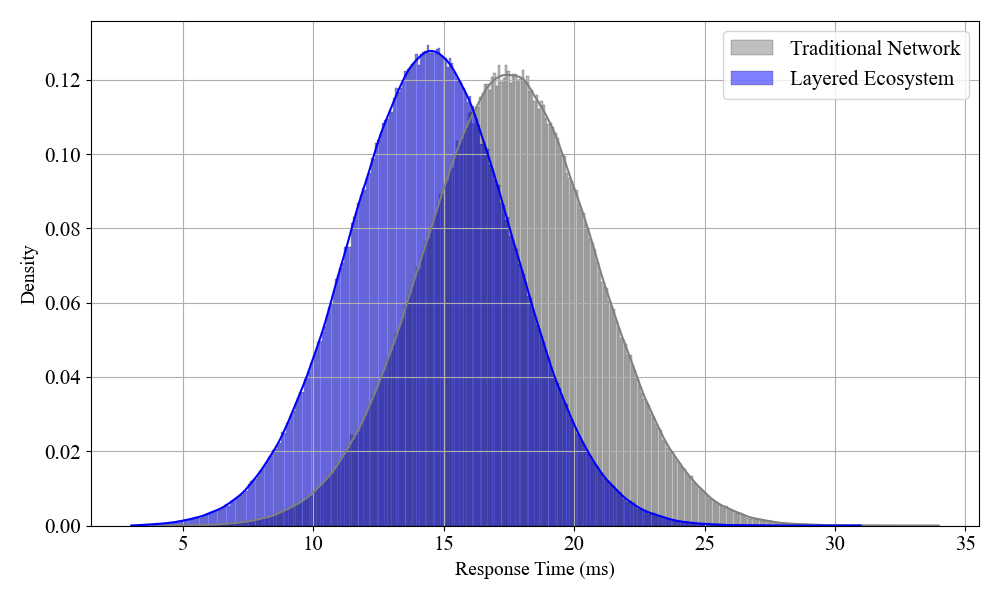}
		\label{fig:5-1}
	}
	\hspace{0.01\textwidth}
	\subfigure[Trends in edge autonomy and population coordination.]{
		\includegraphics[width=0.4\textwidth]{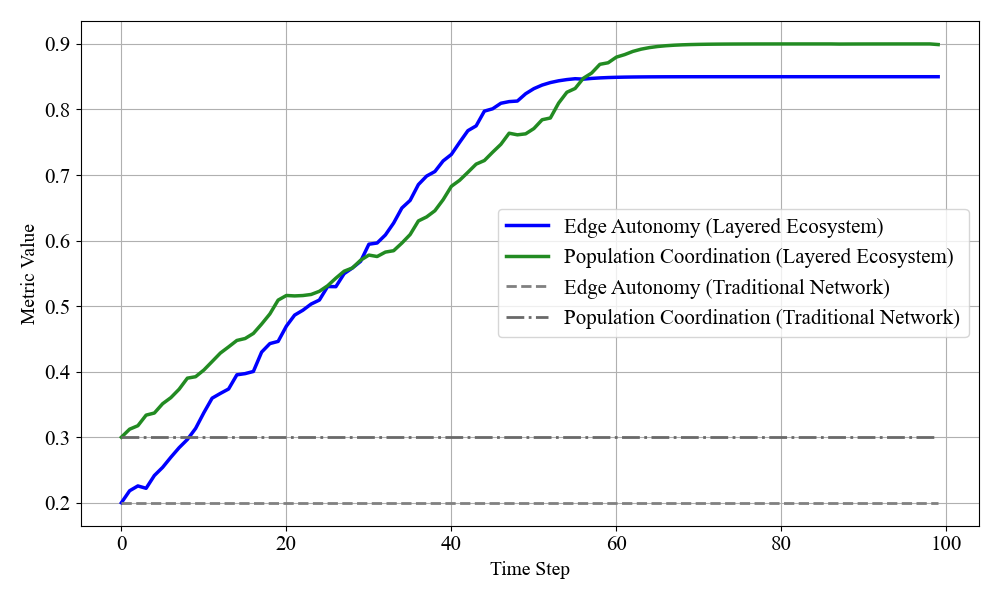}
		\label{fig:5-2}
	}
	\caption{Comparative results of showcase.}
	\label{fig:5-combined}
\end{figure*}

In the first three steps, we constructed a multi-layer DT system from an engineering perspective, tailored to future network architectures. This system integrates real-time sensing at the local layer, dynamic organization of edge populations, and long-term governance at the cloud. Together, these elements form a self-adaptive and self-evolving digital ecosystem, as illustrated in Figure~\ref{fig:3}. Throughout the design process, we drew from natural systems, introducing mechanisms such as population coordination, functional differentiation, information propagation, and feedback control to support dynamic stability similar to ecological systems.

As the system moves into long-term operation, its internal behavior will begin to exhibit social patterns beyond traditional engineering control. For example, cooperation and role division among DTs within edge populations resemble organizational structures seen in social groups. Strategy delivery, resource scheduling, and task competition also show characteristics of incomplete-information games. At this stage, theories from governance studies, behavioral science, and trust modeling can offer tools to improve coordination, resilience, and fairness.

To support sustainable evolution, social science concepts should be embedded into system modeling. Utility functions based on behavioral economics can guide decision-making under competition. Organizational sociology can support local negotiation and coordination among edge DTs. Institutional economics and digital ethics help assess the fairness and acceptability of cloud-level strategies and enforcement policies.

In practical terms, our approach originates from real communication systems and is then enhanced through insights from the social sciences. The DT environment allows large-scale simulation and evaluation beyond physical communication network constraints, helping to identify effective strategies in a virtual setting. These proven strategies are then gradually applied back to original network, guiding real-world protocols and resource policies. In this way, the digital ecosystem grows from reality, surpasses it through digitized experimentation, and finally returns to reality to improve it — forming a healthy cycle for a digital society.

\section*{Showcase: Ecosystem for Smart City Traffic}

To evaluate performance in both task responsiveness and ecological governance, we focus on traffic coordination in a simple smart city using a multi-layer DT system composed of local, edge, and cloud layers. The urban setting includes six RSUs deployed at key intersections, each managing approximately 200 vehicles equipped with onboard devices.  Each vehicle maintains a lightweight local DT that continuously reports its location, speed, energy status, and navigation intent. RSUs collect data from their respective vehicles to form edge-level digital populations and  the cloud DT aggregates summaries from all edges for long-term strategy optimization. 

In terms of task responsiveness, simulation results in Fig.~\ref{fig:5-1} reveal the significant advantages of the proposed hierarchical autonomous driving ecosystem. The distribution of response times shows that vehicles operating under the layered architecture consistently exhibit lower latency and tighter variance compared to those relying on conventional cloud-centric approaches. This improvement stems from the system's ability to handle most real-time demands at the local individuals or edge population.

From the perspective of ecological governance, Fig.~\ref{fig:5-2} demonstrates that the layered system promotes a gradual enhancement of edge autonomy and inter-population coordination. Unlike traditional networks where centralized control leads to static and constrained ecological indicators, the evolving autonomous driving population exhibits growing local decision-making capacity and cooperative behavior across RSUs. This autonomy and self-organization capability are further reinforced by top-down strategy updates from the cloud based on the global historical patterns and local traffic conditions.

Overall, the comparative results highlight the dual advantages of the proposed architecture: the multi-layer digital twin system not only accelerates response speed but also fosters distributed intelligence through ecological evolution, demonstrating its potential as an adaptive ecological framework capable of long-term governance.

\section*{Conclusion}

This paper presents a multi-layer DT architecture that organizes communication networks into an evolving digital ecosystem. Local devices are represented as individual DTs, coordinated at the edge layer into dynamic digital populations, and guided by long-term strategies at the cloud level. Drawing from natural principles such as population coordination, functional differentiation, and feedback-based adaptation, the system is designed to support decentralized intelligence, scalable collaboration, and continuous learning across layers.

As the ecosystem enters long-term operation, its complexity and autonomy demand a broader governance perspective. We propose the integration of social science principles to support sustained balance and ethical alignment. Concepts such as reputation, cooperation, and institutional trust can be encoded into system protocols to guide behavior and resolve conflicts. Through this integration, the DT-based network evolves not only as a technical system but also as a self-regulating digital society.

\section*{Author contributions}

G.Z. conceived the main idea of the layered ecosystem and conducted the theoretical modeling and simulation experiments. He also wrote the initial draft of the manuscript. D.K. supervised the overall research, provided strategy guidance on the system architecture and governance framework, and contributed to manuscript revision. Both authors discussed the results and approved the final version of the manuscript.

\bibliographystyle{IEEEtran}  
\bibliography{ref}            

\end{CJK}

\end{document}